\documentclass[a4paper,fleqn,usenatbib]{mnras}

\usepackage{newtxtext,newtxmath}

\usepackage[T1]{fontenc}
\usepackage{ae,aecompl}

\usepackage{graphicx}	
\usepackage{amsmath}	
\usepackage{amssymb}	
\usepackage{booktabs}
\usepackage{multirow}
\usepackage{xcolor} 

\title[Post-CE binaries with He-driven stable RLOF]{Post-common envelope binary systems experiencing helium-shell driven stable mass transfer}

\author[G.~M.~Halabi, R.~G.~Izzard and C.~A.~Tout]{
Ghina M. Halabi,$^{1}$\thanks{E-mail: gmh@ast.cam.ac.uk}
Robert G. Izzard$^{1,2}$
and Christopher A. Tout$^{1,3}$
\\
$^{1}$
Institute of Astronomy, Madingley Road, Cambridge, CB3 0HA\\
$^{2}$Astrophysics Research Group, Faculty of Engineering and Physical Sciences, University of Surrey, Guildford, GU2 7XH\\
$^{3}$Monash Centre for Astrophysics, School of Physics and Astronomy, Monash University, Clayton VIC 3800, Australia
}

\date{Accepted XXX. Received YYY; in original form ZZZ}

\pubyear{2015}

\begin{document}
\label{firstpage}
\pagerange{\pageref{firstpage}--\pageref{lastpage}}
\maketitle

\begin{abstract}
We evolve stellar models to study the common envelope (CE) interaction of an early asymptotic giant branch star of initial mass $5\,\rm M_{\odot}$ with a companion star of mass ranging from $0.1$ 
to $2\,\rm M_{\odot}$. We model the CE as a fast stripping phase in which the primary experiences rapid mass loss and loses about 80 per cent of its mass. The post-CE remnant is then allowed to thermally readjust during a Roche-lobe overflow (RLOF) phase and the final binary system and its orbital period are investigated. We find that the post-CE RLOF phase is long enough to allow nuclear burning to proceed in the helium shell. By the end of this phase, the donor is stripped of both its hydrogen and helium and ends up as carbon-oxygen white dwarf of mass about $0.8\,\rm M_{\odot}$. We study the sensitivity of our results to initial conditions of different companion masses and orbital separations at which the stripping phase begins. We find that the companion mass affects the final binary separation and that helium-shell burning causes the star to refill its Roche lobe leading to post-CE RLOF. Our results show that double mass transfer in such a binary interaction is able to strip the helium and hydrogen layers from the donor star without the need for any special conditions or fine tuning of the binary parameters.

\end{abstract}




\section{Introduction}\label{sec:intro}
Mass transfer is a critical feature of the evolution of close binary systems. This direct interaction between the stellar components has key implications to all stages of stellar evolution and distinguishes binary evolution from that of single stars. The rate of this mass transfer determines the fate of the remnants such as Algols, X-ray binaries, contact binaries, cataclysmic variables and double-degenerate systems involved \citep{PW85,DE17}.\\
\indent In a frame rotating with a tidally locked, circular binary system, the effective gravitational potential is an equipotential surface through the inner Lagrangian point that defines the Roche lobe of each star. The volume enclosed by the Roche lobe determines the Roche lobe radius of each star \citep{Eg83}. If either star fills its Roche lobe then material overflows from its outer layers through the inner Lagrangian point that connects the two Roche lobes where the gradient of the effective potential vanishes. Stable mass transfer occurs by Roche lobe overflow (RLOF) by virtue of either the slow expansion of the star because of nuclear evolution or of orbital contraction by angular momentum losses from gravitational radiation, magnetic braking in stellar winds or tides if the Roche-lobe filling star must be spun up.\\ \indent
Some or all of the transferred material may be captured by the companion and consequently the evolution of both the donor and the accretor is expected to differ from that of similar single stars. Binary systems that have long orbital periods allow the more massive star to reach the red giant phase before filling its Roche lobe. The giant star then has a deep convective envelope and runaway mass transfer reaches dynamical time-scales \citep{We84,Iv13a}. This also happens if the Roche lobe-filling star is significantly more massive than its, most often, main-sequence companion \citep{Pa65}.\\ 
\indent Because of its relatively long thermal time-scale, the accreting star cannot capture all the material transferred from the donor star, so material accumulates in a common envelope (CE) surrounding both stars leading to the formation of a CE system \citep{Pa76}. As the dense companion plunges into the giant's envelope, gravitational drag forces cause the orbit of the embedded binary to shrink dramatically and the core of the donor and its companion star spiral inward through their common envelope \citep{LS88, TS00,Pa12a}. Possible outcomes include the release of sufficient energy to drive off the entire envelope as the giant core and MS star spiral in, resulting in a closer binary, or merging of the stars. This explains the observed short-period degenerate systems such as cataclysmic variables, close binary pulsars and close double white dwarf binaries which, otherwise, cannot be explained by angular momentum losses by gravitational waves or magnetic winds \citep{IL93}.\\
\indent There are several variations of the treatment of the CE and many studies have been carried out \citep[see][for a review]{Iv13a}. Most rely on analytical prescriptions based on energetic considerations \citep{We84,IT85} where the efficiency of the conversion of orbital energy of the binary into kinetic energy of the outflow is assumed. Another prescription based on angular momentum considerations \citep{Ne00,NT05} parametrizes the angular momentum of the ejected envelope. However, this has been found to be less useful than the energy budget approach for predicting the outcome of the CE and constraining the parameters of the possible progenitors of observed systems \citep{ZO10}. Other approaches include a more accurate description of the ejection conditions, such as the donor star's structural response to adiabatic mass loss \citep{DT10}. However, the efficiency of the ejection process remains uncertain.\\
\indent A standard treatment of the CE is the energy formalism \citep{We84} in which the final separation of the binary is determined by relating the loss in the orbital energy of the system to the binding energy of the released envelope. A large fraction of the orbital energy released in the spiral-in process is transferred into the expansion of the envelope with efficiency $\alpha_{\rm CE}$ \citep{LS84}. The envelope is then ejected when the total deposited orbital energy, $\Delta E_{\rm orb}$, exceeds the
binding energy of the envelope, $E_{\rm env}$, or
\begin{equation}
\alpha_{\rm CE}\,\Delta E_{\rm orb} \geq E_{\rm env},\label{eqn:CE}
\end{equation}
and
\begin{equation}
\Delta E_{\rm orb} = \frac{GM_{\rm c1}M_{2}}{2 a_{\rm f}} - \frac{M_{\rm c1}M_{2}}{2 a_{\rm i}},  \label{eqn:CE2}
\end{equation}
where $M_{1}$ and $M_{\rm c1}$ are the masses of the giant and its core, respectively, $M_{2}$ is the mass of the secondary, which is not affected, and $a_{\rm i}$ and $a_{\rm f}$ are the initial and final separations, before and after the common envelope, respectively \citep{Hu02}.\\
\indent The efficiency parameter $\alpha_{\textnormal{CE}}$ and the density profile of the envelope determine the final separation of the system. Because the CE phase involves various complex physical processes occurring on very different time-scales, $\alpha_{\rm CE}$ cannot yet be determined from first principles and it thus constitutes a simple prescription for the complex hydrodynamical interaction taking place during and after the spiral-in phase.
Moreover, $\alpha_{\textnormal{CE}}$ is probably not a constant \citep{RT95, DE11b, DA11} but is often set to $\alpha_{\textnormal{CE}}=1$ \citep{Hu02}. Some studies attempt to constrain $\alpha_{\textnormal{CE}}$ with certain systems and then assume it is the same for all similar systems. For example, \citet{BR01} study low-mass black-hole X-ray binaries (soft X-ray transients) with main-sequence companions that have formed through case~C mass transfer and constrain $\alpha_{\rm CE}$ to be $0.2$ to $0.5$. \citet{RT95}, on the other hand, model the magnetic dynamo owing to differential rotation within the envelope. They find that $\alpha_{\rm CE}$ lies in the range 0.5 to 1.0 but it depends on the initial state of the envelope and changes during the evolution. Following \citet{RT95}, later work by \citet{TO97} favours $\alpha_{\rm CE}=1$. Therefore the energy formalism is useful to predict the fate of CE evolution but its outcome is not fully understood. Multi-dimensional hydrodynamical simulations that model the CE evolution \citep{Pa12a, RT12} cannot be used to relate the pre- and post-CE configurations because they end after a rapid spiral-in phase before most of the envelope is unbound. For these reasons, the common envelope phase is one of the most uncertain processes in binary stellar evolution \citep{Iv13a} and realistic self-consistent models are still lacking. This affects our understanding of the evolution of close binary systems such as compact X-ray binaries, cataclysmic variables, merging gravitational wave sources and Type~Ia~supernovae.\\
\indent Evidence for CE evolution is provided by plenty of observed systems, such as cataclysmic variables (CVs) and double-degenerate binaries. Close binary systems containing a carbon-oxygen white dwarf and a main-sequence star with periods of one day or less, including CVs, are well known \citep{Kn11, Ri12}. These can only be explained if a significant amount of mass and angular momentum are removed from their precursor system. Other possible examples of CE events are planetary nebulae (PNe) with a close binary at their centre \citep{BO78,JB17}. A recent observational study of optical spectra of a large sample of Galactic planetary nebulae by \citet{WE18} shows features such as hydrogen-deficiency or stellar lines that are shifted with respect to the nebular ones for example, which suggest a binary core in several systems. The connection between duplicity and the observed nebular structure has been proposed on theoretical and observational grounds. Theoretically, it was predicted that some, perhaps even all, PNe should be the outcome of a CE \citep{DE09}. An AGB star in a binary system overflows its Roche lobe and interacts with its companion unless the system is very wide. This leads to a CE, a spiral-in of the companion and a tight final orbit of a few hours to a few days. Aspherical PNe with bipolar ejecta featuring dense equatorial rings and higher-velocity polar jets are thought to be the products of binary interactions \citep{We08,DE11a}. Observational evidence for this connection is the significant change in the radii of the secondary stars in planetary nebulae with extremely close binary nuclei. These companions are reported to have larger radii than expected for main-sequence stars of the same masses \citep{OB01,AI08}. Although it is uncertain whether the mass of the secondary is substantially affected during the CE phase \citep{PL85,SA98}, the observed oversized secondary companions are thought to have either recently emerged from a CE, and hence are out of thermal equilibrium, or their mass and radius changed because of mass transfer during the CE phase, perhaps even both. On the other hand, the curious emerging class of optical transients with predominantly red spectra observed in the local Universe and commonly dubbed as luminous red novae or intermediate luminosity optical transients \citep{Mar99,Bla17} are perhaps the best CE candidates observed so far and can thus be used to measure CE outburst energies and durations. While an agreement between some of their features and model predictions has been reported \citep{Iv13b}, the field of CE hydrodynamics and associated radiative transfer remains an area of active research \citep{GA17}.\\
\indent Currently, neither observations nor theory provide strong constraints on the stellar evolution during or immediately after the CE phase. Numerical simulations of CE evolution \citep{RT12,Pa12a} including only gravitational drag tend to show the companion star rapidly spiralling into the envelope of the giant as angular momentum is lost by the orbit. These simulations start with the companion already at the surface of the giant. When begun at the onset of Roche lobe overflow \citep{IA17}, the establishment of the common envelope begins slowly but once in place the same rapid inspiral of the cores is seen. At the end of this phase the envelope has expanded but remains bound. It is what follows that we model. Without evidence to the contrary we suppose that the envelope is removed by a super wind, similar to the strongest winds observed from AGB stars \citep{VW93} on a time-scale of a few thousand years or so. This has also been proposed by \citet{HG18}, who suggest that the envelope is lost by dust-driven winds following the CE event similar to processes operating in the ejection of the envelopes of AGB stars. We consider a binary system in which the more massive star fills its Roche lobe at the early asymptotic giant branch phase (EAGB). An EAGB star has completed core helium burning and is characterized by a core essentially consisting of carbon and oxygen, the main products of helium-burning, surrounded by a helium-burning shell and a hydrogen-burning shell, which is the main energy source in the giant star. CE evolution with an EAGB star must be common and the EAGB structure  makes them interesting objects if stripped. They are expected to evolve to hybrid white dwarfs (low-mass carbon-oxygen cores with thick helium envelopes) and they may sustain nuclear burning after the CE phase as we show in Section~\ref{sec:WD}. On the EAGB, a star expands to larger dimensions than on the RGB. Thus when it expands to a radius $R_{\rm{CE}}$, which exceeds the maximum radius reached on the RGB, it can undergo case~C mass transfer \citep{Ki67}. We strip the star by applying fast mass loss to mimic a CE event. Once the system detaches, we allow the donor to thermally adjust and refill its Roche lobe. We choose the mass of the companion such that the subsequent RLOF is stable and study the behaviour of the binary system.\\
\indent \citet{NO94} use such double mass-transfer events to model the evolution of the progenitors of Type~Ic~supernovae and suggested this as a possible evolutionary scenario for hypernovae. \citet{CR07} consider a binary orbit that allows interaction between the star and its companion but not so close as to merge. This binary interaction removes only the hydrogen envelope of the progenitor star and subsequent shedding of the helium-rich layer occurs by strong radiatively driven winds. \citet{NO01} point out that the helium layer may be removed with a second mass-transfer event given the right conditions of initial mass and separation. This conclusion is based on earlier work by \citet{NO95} where they assume that the first mass transfer occurs when the primary has formed a helium core (case B mass transfer). They argue that this is possible in low-mass helium stars which have large enough radii to fill their Roche lobes. Larger-mass helium stars, on the other hand, have radii too small to fill their Roche lobes as seen in results by \citet{HA86}, for example. Suggested explanations for why they remain small and hot can be found in \citet{Egg06}. These larger-mass helium stars, however, have large enough luminosities to lose most of their helium layer by strong winds instead. We discuss the sensitivity of the removal of the hydrogen and helium layers to the initial conditions in Section~\ref{ssec:comp}.\\ 
\indent We focus on a scenario in which, after the CE event, the binary system ejects its envelope and avoids merging. This determines the chosen post-CE separation of our binary. We also assume that its main-sequence companion does not fill its Roche lobe. However, because it is unclear how stars behave during the extremely rapid, possibly adiabatic, mass loss of the CE phase \citep{Iv13a} and how their radii are affected by CE evolution, we investigate various evolutionary sequences with different post-CE orbital separations and study the effect on the final state of the remnants. We also investigate changing the companion mass on the fate of the resulting binary system. In Section \ref{sec:evol} we present our evolutionary code and the evolution of our model through CE and RLOF. The dependence of our model on the initial conditions is discussed in Section \ref{sec:Res}. We conclude in Section~\ref{sec:conc}.
\section{Evolutionary Models}\label{sec:evol}
We use the version of the Cambridge stellar evolution code
\textsc{stars}\footnote{The code is publicly available at \url{http://www.ast.cam.ac.uk/~stars}} described by \citet{ST09}. The code was originally written by \citet{Eg71,Eg72} and updated by \citet{PO95}. Evolutionary model sequences are produced by solution of a set of discretized, one-dimensional, quasi-static stellar-structure equations. The meshpoints are distributed in a non-Lagrangian mesh \citep{Eg71}. We set the convective mixing-length \citep{BV58} parameter $\alpha_{\rm MLT} = 2$ and we assume a metallicity $Z = 0.02$. The input physics is described by \citet{HT14}.
\subsection{Mass loss during common envelope evolution}
\label{sec:ML}
To illustrate CE formation and evolution, we start by considering a binary system with a primary star of initial mass $5\,\rm M_{\odot}$. The system is in a circular orbit with a sufficiently long orbital separation that the more massive component evolves to the EAGB, $0.12\,\rm Gyr$ after it evolved off the ZAMS, before filling its Roche lobe. The EAGB star has a carbon-oxygen core of mass $0.53\,\rm M_{\odot}$, a helium layer of mass $0.47\,\rm M_{\odot}$ and a hydrogen envelope of $4\,\rm M_{\odot}$. It fills its Roche lobe and starts stripping when it reaches a defined radius on the EAGB of $R_{\rm CE}=100\,\rm R_{\odot}$. There is no mass transfer prior to the EAGB.\\
\begin{figure}
\includegraphics[scale=0.32]{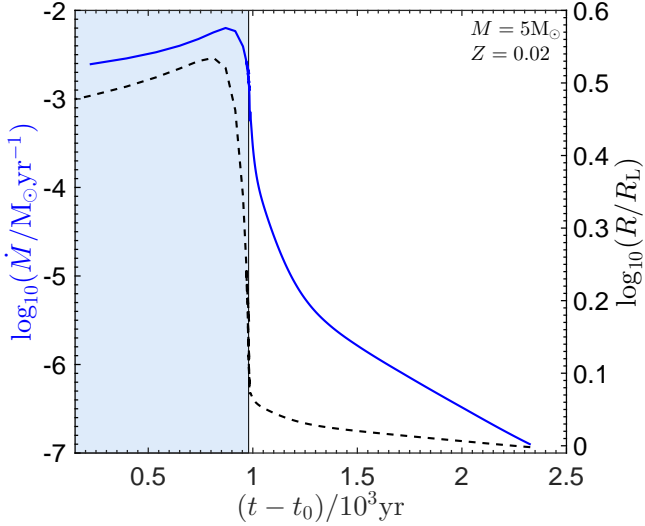}
    \caption{Mass-loss rate from the $5\,\rm M_\odot$ star (solid blue line) and radius (dashed black line), normalized to the Roche lobe radius, as a function of time. Mass loss starts at time $t_{0}$. A high mass-loss rate is applied during the fast stripping phase (shaded region) and we switch to an exponentially decaying rate as the radius approaches $R_{\rm L}$ to ensure a smooth transition between the two.}
    \label{fig:ML}
\end{figure}
\indent We apply a fast mass-loss rate, $\dot{M}_{\rm CE}$, to mimic common envelope dynamical mass transfer. Because we need a large mass-loss rate that drops off at small radii, we model the stripping phase with a \citet{Re75} mass-loss rate with a large multiplier $\eta$,
\begin{equation}
\dot{M}_{\rm CE} = -\eta\,\frac{R}{\rm R_{\odot}}\frac{L}{\rm L_{\odot}}\frac{\rm M_{\odot}}{M} \,\rm M_{\odot}\,\rm yr^{-1},\label{eqn:ML-Reim}
\end{equation}
where $\eta = 2 \times 10^{4}$, the highest that ensures model convergence, and $L$, $M$ and $R$ are the stellar luminosity, mass and radius, respectively.
We end the stripping phase when the radius, $R_{1}$, of the naked helium star approaches an arbitrarily chosen post-CE Roche lobe radius, $R_{\rm L}=24.5\,\rm  R_{\odot}$ that ensures all hydrogen is stripped off. Because this final radius is uncertain, we investigate the sensitivity of our results to its choice in Section \ref{ssec:comp}. To avoid numerical artefacts owing to a sharp cut-off in the mass-loss rate when $R_{1}=R_{\rm L}$, we employ an exponentially declining mass-loss rate,
\begin{equation}
|\dot{M}|=\textnormal{min}\Big\{{|\dot{M}_{\rm CE}|\, ,\, A\exp\big[C (R-  R_{\rm L})\big]}\Big\},\label{eqn:ML}
\end{equation}
where $\dot{M}_{\rm CE}$ is given by equation~(\ref{eqn:ML-Reim}), $A=10^{-14}$ M$_{\odot}\,\rm yr^{-1}$ and $C=50/\rm R_{\odot}$. This expression and choice of parameters ensure that when $R_{1}> R_{\rm L}$ the mass-loss rate is fast $\dot{M}_{\rm CE}$ while, as $R_{1}$ approaches $R_{\rm L}$ from above, the mass-loss rate decays exponentially. It also ensures that mass transfer is stable on a nuclear or thermal time-scale when $R_{1} \approx R_{\rm L}$. Note that the parameter $C$ controls how fast the exponential rate drops. It is chosen such that it is high enough to ensure rapid mass loss but without causing a sharp transition between the fast and the exponential rates. A very large $C$ causes the system to oscillate between the two rates and become unstable even with a shorter time-step, while a smaller $C$ does not ensure a fast enough mass-loss rate as would be expected in a CE event.\\   
\indent Fig.~\ref{fig:ML} shows the mass-loss rate, $|\dot{M}|$, during the rapid stripping phase. It reaches a maximum of $6\times 10^{-3}\,\rm M_\odot\, yr^{-1}$ and lasts for $2.3 \times 10^{3}\,\rm yr$. About $4\,\rm M_\odot$ is lost by the primary and its remnant, the stripped core, is reduced to $M_{1}=0.997\,\rm M_\odot\,$.\\ 
\indent Several estimates exist for the duration of the CE phase. \citet{Po01}, using a stellar evolution code, predicts that a CE phase may last $100$ to $1000\,\rm yr$. CE 3D hydrodynamic simulations of the dynamical in-fall phase (not including ejection) by \citet{RT12} estimate it to be longer than about $50\,\rm d$. \citet{Pa12a} find that most of the in-spiral happens within $200$ to $300\,\rm d$, and \citet{IN16} find this to be a few hundred days. The lack of conclusive observational evidence leaves the CE duration unconstrained and motivates the search for observable signatures of CE evolution which could serve as diagnostic of the instabilities of the spiral-in and the history of the mass loss associated with CE evolution.\\
\indent The donor, in our case, is stripped of most of its envelope during the CE phase over $2.3 \times 10^{3}\,\rm yr$. The mass of the helium layer is $0.47\,\rm M_\odot$ and the surface is comprised of a thin hydrogen layer of $3\times 10^{-2}\,\rm M_\odot\,$, as expected after a dynamical CE phase. It would be expected to look like a post-AGB star.   
\subsection{Roche-lobe overflow} \label{ssec:RLOF}
At the end of the CE phase, the stripped core of mass $0.997\,\rm M_{\odot}$ is within its Roche lobe and the system detaches. Upon any subsequent mass transfer, we switch to conservative RLOF. We use a relation between the mass transfer rate and the excess radius of the form calculated by \citet{Je69} and published by \citet{PS72},
\begin{equation}
\left|\dot{M}_{\textnormal{RLOF}}\right|= \Gamma \big[(R_{1}-R_{\rm L})/R\big]^{3},
\end{equation}
where $\Gamma$ is chosen to be sufficiently large to ensure that the radius adjusts and remains close to $R_{\rm L}$ during mass transfer. We have chosen the mass of a companion such that the mass ratio ensures any further mass transfer is stable. The post-CE system has $ M_{1}=0.997\,\rm M_{\odot}$, $M_{2}=0.78\,\rm M_{\odot}$ with a period of $49.3\,\rm d$.\\
\begin{figure}
	\includegraphics[scale=0.32]{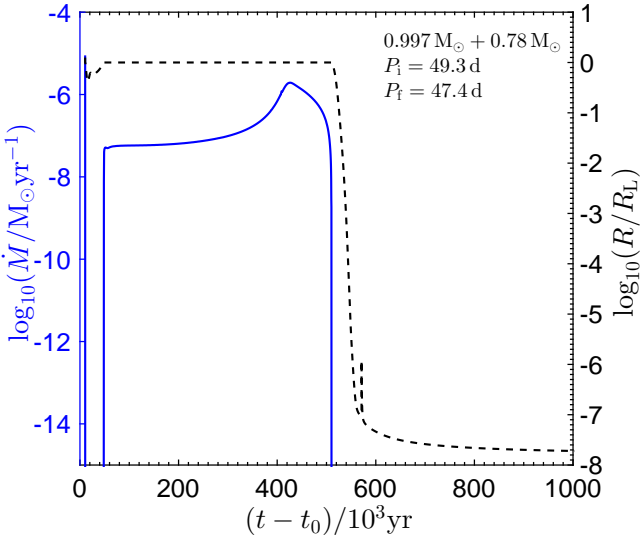}
    \caption{Subsequent evolution of the donor after the CE phase. The RLOF mass loss, $\rm M_{\odot}$, is shown in blue as a function of the age $t$ (post-CE RLOF). It shows a fast drop-off at the beginning, which is just when the CE ends. Time $t_{0}$ is when mass loss starts (at the beginning of the CE phase, as Fig.~\ref{fig:ML}). The radius of the primary, normalized to $R_{\rm L}$, is shown by a dashed black line. When the star is within its Roche lobe, the mass loss stops and it restarts when the star expands. The spike in the radius after RLOF is due to a helium-shell flash at the surface, similar to hydrogen-shell flashes on the surface of white dwarfs.}
    \label{fig:ML_R}
\end{figure}
\begin{figure}
\includegraphics[scale=0.32]{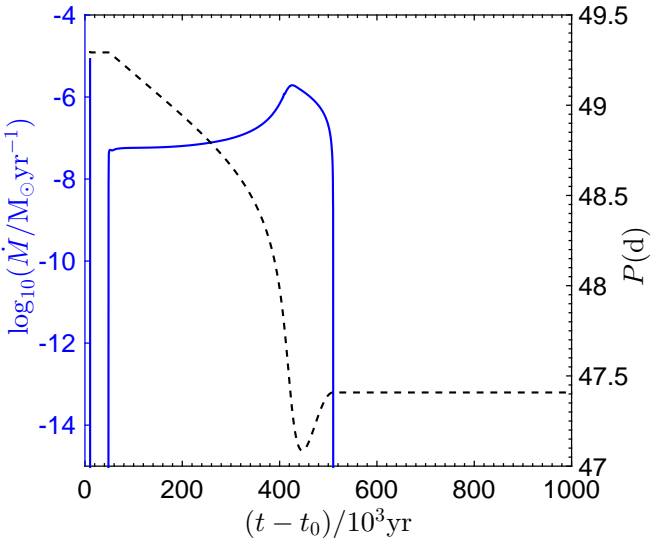}
    \caption{The evolution of the orbital period, P, (dashed black line) in time $t$ during the post-CE RLOF phase. Time $t_{0}$ is when mass loss starts (at the beginning of the CE phase, as Fig.~\ref{fig:ML}). The mass-loss rate is reproduced as a solid blue line. The period $P$ falls until $M_{1}=M_{2}$ then the system expands again.}
    \label{fig:ML_P}
\end{figure}
\indent Because we set $R_{1}=R_{\rm L}=24.5\,\rm  R_{\odot}$ at the end of the CE, this then fixes the orbital separation, $a$, given $M_{2}$. We calculate $a$ from the Roche lobe radius of the primary star using the expression of \citet{Eg83},
\begin{equation} 
\frac{R_{\rm L}}{a}=\frac{0.49q^{2/3}}{0.6q^{2/3}+\ln(1+q^{1/3})},
\label{eqn:Egg}
\end{equation}
where $q=M_{1}/M_{2}$ and we find the period using Kepler's third law.\\
\indent Fig.~\ref{fig:ML_R} shows the evolution of $|\dot{M}_{\textnormal{RLOF}}|$ together with the radius, normalized to the star's Roche lobe, after the CE is ejected. 
The star is within its Roche lobe for about $3.8\,\times 10^{4}\,\rm yr$, during which there is no mass transfer. The star then expands owing to its nuclear evolution as a helium-burning star, fills its Roche lobe again and mass transfer restarts. The rate $|\dot{M}_{\textnormal{RLOF}}|$ reaches $10^{-7}$ to $10^{-6}\,\rm M_{\odot}\, yr^{-1}$ until eventually $R_{1}<R_{\rm L}$ after about $5\times 10^{5}\,\rm yr$.\\
\indent Fig.~\ref{fig:ML_P} illustrates the change in the orbital period during this phase, when the stripped post-CE core loses $0.15\,\rm M_{\odot}$. This stable mass transfer is quite prolonged and planetary nebulae have a lifetime of about $10^{4}\,\rm yr$, an estimate that is weakly dependent on the mass of the central star \citep{JA13}. If post-EAGB stars make such nebulae, their central binaries are likely to be undergoing stable mass transfer and this should be observable.\\
\indent Because mass is transferred from the more massive star the orbit shrinks. The minimum separation is when $M_{1}=M_{2}$, after which the mass ratio inverts, mass loss slows and the system detaches soon afterwards. The same separation is then maintained until the primary evolves into a white dwarf of mass $ M_{1}=0.847\,\rm M_{\odot}$. The secondary has mass $ M_{2}=0.93\,\rm M_{\odot}$, and by the end of the post-CE RLOF phase the final separation is $67\,\rm R_{\odot}$ and the period is 47.4\,d.
%
\section{Results and Discussion}\label{sec:Res}
We investigate the composition profile of the remnant after the CE and the subsequent RLOF phase. We also study the sensitivity of these results to varying the mass of the secondary and the initial separation at which the CE begins. 
\subsection{White dwarf composition}\label{sec:WD}
\begin{figure}
	\includegraphics[scale=0.32]{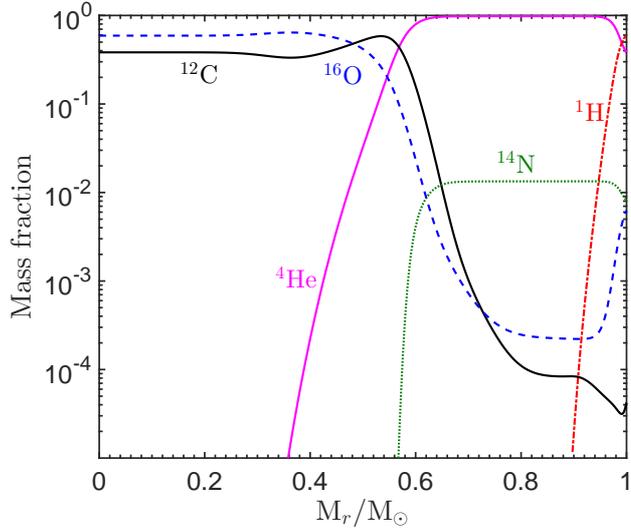}
    \caption{Composition profile, as a function of mass coordinate $M_{r}$, of the interior of the $0.997\,\rm M_{\odot}$ remnant after the CE and before the RLOF phase. The abundance profiles correspond to $^{1}$H (dot-dashed red line), $^{4}$He (solid magenta line), $^{12}$C (solid black line), $^{14}$N (dotted green line)  and $^{16}$O (dashed blue line).}
    \label{fig:WD-RLOF}
\end{figure}

\begin{figure}
	\includegraphics[scale=0.32]{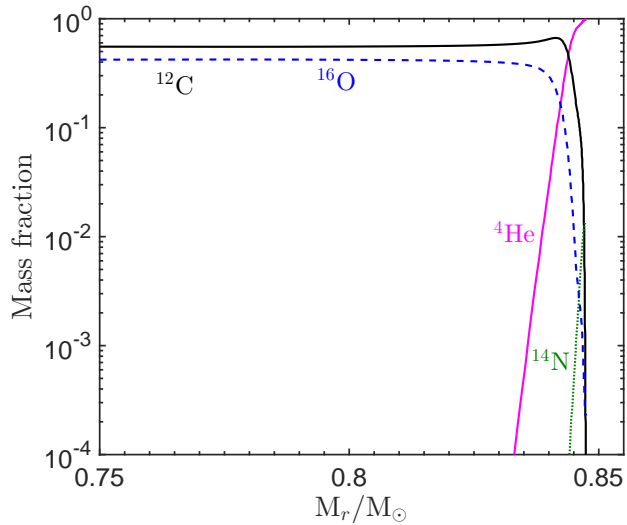}
    \caption{The surface abundance profiles, as a function of mass coordinate $M_{r}$, of $^{4}$He (solid magenta line), $^{12}$C (solid black line), $^{14}$N (dotted green line) and $^{16}$O (dashed blue line) in the $0.847\,\rm M_{\odot}$ stripped core after the post-CE RLOF phase. The surface helium layer has a mass $5.8\times10^{-3}\,\rm M_{\odot}$.}
    \label{fig:WD}
\end{figure}

Figs.~\ref{fig:WD-RLOF} and \ref{fig:WD} show the donor's internal composition after the CE and the RLOF phase, respectively. During the prolonged post-CE RLOF, the mass of the carbon-oxygen core grows by helium shell burning from about $0.530\,\rm M_\odot$ to about $0.846\,\rm M_\odot$. It has central abundances $X_{\rm C} = 0.38$ and $X_{\rm O} = 0.59$. All the surface hydrogen is stripped from the remnant and it has a thin helium layer on the surface, of mass $5.8\times10^{-3}\,\rm M_{\odot}$. Because the surface helium abundance is $X_{\rm He} = 0.98$, this would appear as an extremely helium rich subdwarf. CE evolution has been proposed as a possible evolutionary path for an observed spectroscopic binary with a helium-rich subdwarf component rather than a merge \citep{NA12}.\\
\indent The material accreted by the secondary during the post-CE RLOF phase changes in composition from hydrogen and helium rich with traces of carbon and oxygen depleting the surface composition of the donor (Fig.~\ref{fig:WD-RLOF}) to predominantly helium at the end of the post-CE RLOF phase as depicted in the final surface abundance profile of the donor in Fig.~\ref{fig:WD}. If the secondary is a white dwarf accreting this helium-rich material at the predicted $\dot{M}_{\textnormal{RLOF}}$ rates of $10^{-7}$ to $10^{-6}\,\rm M_\odot\, yr^{-1}$, it is expected to burn helium into carbon and oxygen stably because the surface degeneracy is raised \citep{No82}. The system may also be observed as a supersoft X-ray source \citep{VA92, Di97} if the accretion is fast enough to sustain fusion on its surface. Such systems with helium-rich donors have also been found to be the dominant single degenerate channel for Type~Ia~supernova with the shortest delay times \citet{CL14}. If the accreting secondary is a main-sequence star of mass $0.78\,\rm M_{\odot}$, its Kelvin-Helmholtz time-scale is about $40\rm \,\rm Myr$ \citep{Hu02}. The RLOF lasts for about $5\times 10^{5}\,\rm yr$, during which it accretes about $0.15\,\rm M_{\odot}$. We calculate how this accretion affects the secondary and find that for a $0.78\,\rm M_{\odot}$ of radius $0.79\,\rm R_{\odot}$, accreting at this rate swells it up to $0.84\,\rm M_{\odot}$. Given that the separation of the binary drops from about $68\,\rm R_{\odot}$ to $66\,\rm R_{\odot}$ during RLOF, the system is wide enough to ensure that this fast accretion is not expected to cause the secondary to fill its Roche lobe.
%
%
%
\begin{table*}
  \begin{center}
    \caption{Properties of the post-CE object and the resulting white dwarf (WD) after RLOF for various Roche lobe radii, $R_{\rm L}$, at the end of CE ejection. Listed quantities are the surface hydrogen (X$_{\textnormal{H}}$) and helium (X$_{\textnormal{He}}$) mass fractions, mass enclosed in the hydrogen ($\Delta$M$_{\textnormal{H}}$) and helium layers ($\Delta$M$_{\textnormal{He}}$), orbital period $P$ after each phase and the total stellar mass M$_{\rm t}$.}
    \label{tab:table1}
    \begin{tabular}{lllllll|llllll}
        &\multicolumn{6}{c}{Post-CE}& \multicolumn{5}{c}{Post-RLOF (WD)} &\\ \hline
        \hline
      $R_{\rm L}$/R$_{\odot}$ & X$_{\textnormal{H}}$ & X$_{\textnormal{He}}$ & $\Delta$M$_{\textnormal{H}}/\rm M_{\odot}$ & $\Delta$M$_{\textnormal{He}}/\rm M_{\odot}$ & P/d & M$_{\rm t}/\rm M_{\odot}$ & X$_{\textnormal{H}}$ & X$_{\textnormal{He}}$ & $\Delta$M$_{\textnormal{H}}/\rm M_{\odot}$ & $\Delta$M$_{\textnormal{He}}/\rm M_{\odot}$ & P/d & M$_{\rm t}/\rm M_{\odot}$\\
      \cmidrule(lr){2-7}\cmidrule(lr){8-13}
      50.0 & 0.63 & 0.34 & 3.2$\times 10^{-2}$ & 0.48 & 79 & 0.997 & 0 & 0.98 &0 &5.8$\times 10^{-3}$ & 76 &0.849\\
      33.0 & 0.60 & 0.38 & 2.9$\times 10^{-2}$ & 0.466 & 77 & 0.997 & 0 & 0.98 &0 &5.8$\times 10^{-3}$ & 74 &0.849\\
      24.5 & 0.59 & 0.38 & 3.0$\times 10^{-2}$ & 0.47 & 49 & 0.997 & 0 & 0.98 &0 &5.8$\times 10^{-3}$ & 47 &0.847\\
      5.0 & 0.54 & 0.44 & 2.6$\times 10^{-2}$ & 0.465 & 4.50 & 0.994 & 0 & 0.98 &0 &7.0$\times 10^{-3}$ & 7.4 &0.841\\
      3.0 & 0.49 & 0.48 & 2.0$\times 10^{-2}$ & 0.46 & 2.10 & 0.991 & 0 & 0.98 &0 &6.9$\times 10^{-3}$ & 2.0 &0.832\\
      2.0 & 0.33 & 0.644 & 1.5$\times 10^{-2}$ & 0.40 & 1.15 & 0.983 & 0 & 0.98 &0 &6.8$\times 10^{-3}$ & 1.9 &0.831\\
      1.9 & 0 & 0.98 & 0 & 0.20 & 1.05 & 0.880 & 0 & 0.98 &0 &6.4$\times 10^{-3}$ & 1.1 &0.829\\
      1.5 & 0 & 0.98 & 0 & 0.07 & 0.80 & 0.827 & 0 & 0.98 &0 &6.6$\times 10^{-3}$ & 0.8 &0.824\\
      1.0 & 0 & 0.98 & 0 & 0.07 & 0.40 & 0.821 & 0 & 0.98 & 0 &7.6$\times 10^{-3}$ & 0.4 &0.817\\
      \hline
    \end{tabular}
  \end{center}
\end{table*}
%
\subsection{Companion mass}
\begin{figure}
	\includegraphics[scale=0.335]{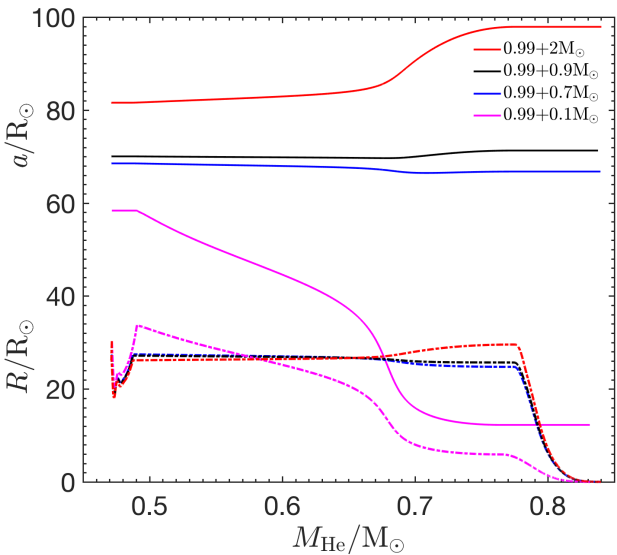}
    \caption{Solid lines show the evolution of the separation $a$ during the post-CE RLOF phase for the various systems under consideration with helium core mass $M_{\rm He}$. The starting point of the lines is the end of the CE phase. The radii of the donors in each system are shown by the dashed lines that have the same colour code as the separations. Note that at the end of the CE $R_{\rm L}=24.5\,\rm R_{\odot}$ in all cases but the orbital separation, $a$, is different by virtue of equation~(\ref{eqn:Egg}).}
    \label{fig:Sep_R}
\end{figure}

We explore the effect the mass of the companion has on the fate of the binary system with secondary masses $M_{2}=0.1$, 0.9 and $2\,\rm M_{\odot}$ in comparison with the system discussed earlier with $M_{2}=0.78\,\rm M_{\odot}$.\\
\indent Fig.~\ref{fig:Sep_R} shows how the orbital separation evolves in the four different systems and how different companion masses result in different ultimate separations even though $R_{\rm L}=24.5\,\rm  R_{\odot}$ at the end of the CE phase is fixed. When $q>1$, the orbit shrinks as mass is transferred from the more massive to the less massive companion during the conservative post-CE RLOF, while for the systems with $q<1$ the orbit expands because the donor is less massive. Also shown in the figure is the radius $R_{1}$ of the donor in each system, colour-coded as the separations. The radius shrinks after the CE phase, causing the mass loss to drop. The primary then expands and fills its Roche lobe so mass loss resumes and the orbit shrinks. Thus the Roche lobe radius becomes smaller causing stable mass transfer by RLOF.\\
\indent The system with $M_{2}=2\,\rm M_{\odot}$ results in a final binary with a relatively long orbital period similar to symbiotic binaries which have periods of a few hundred days. An interesting case arises if the binary has a low mass secondary, as is shown in Fig.~\ref{fig:Sep_R} for the system $0.997+0.1\,\rm M_{\odot}$. This results in a binary with a final orbital separation of about $10\,\rm  R_{\odot}$ and an orbital period of $4.7\,\rm d$. Similar to the $0.997+0.78\,\rm M_{\odot}$ system discussed in Section \ref{sec:WD}, the $0.1\,\rm M_{\odot}$ is expected to remain stable on a thermal time-scale despite the extreme mass ratio, and thus within its Roche lobe at this final orbital separation. Indeed, most observed binary systems which are likely to be post-CE binaries have periods shorter than $10\,\rm d$ \citep{JB17}, and this short-period binary may explain some of them. On the other hand, some have speculated that systems which enter RLOF during the AGB phase may form the shortest-period barium stars \citep{HA95, IZ10}. However, the primary in such systems is a thermally pulsing AGB star, and thus more evolved, and these systems are found to be more generally formed by wind RLOF or wind mass transfer.\\
\indent The lack of known post-CE systems with longer periods may be attributed to observational detection bias against longer period systems \citep{DE08, JB17}. For example, little is known about the evolutionary paths leading to post-AGB binaries with periods $100$ to $1000\,\rm d$ \citep{VW09} and their role in the formation and morphology of PNe.\\
\indent In all our model sequences discussed above, at the end of the post-CE RLOF the donor is stripped of its hydrogen shell and most of its helium shell. It ends up as a CO white dwarf of mass about $0.84\,\rm M_{\odot}$ with a helium surface layer of about $9\times 10^{-3}\,\rm M_{\odot}$. So the double mass transfer that strips away the hydrogen shell and most of the helium shell is not sensitive to the mass of the companion.
\begin{figure}
\includegraphics[scale=0.335]{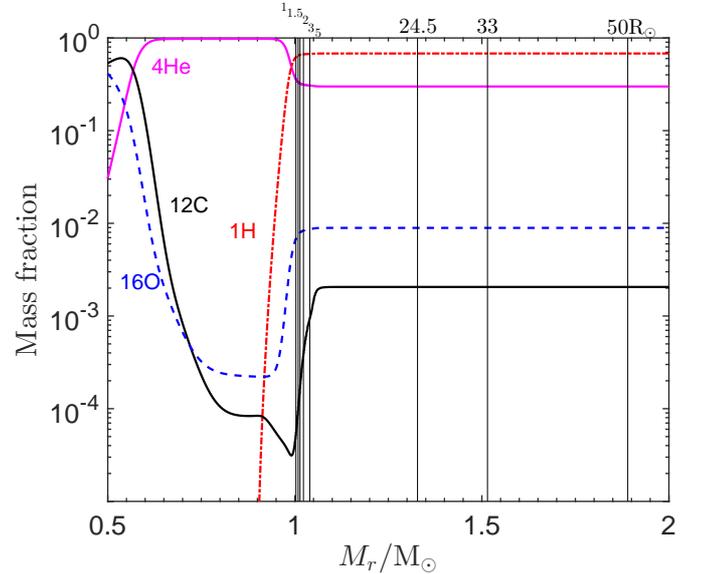}
    \caption{Composition profiles inside the $5\,\rm M_\odot$ star before the CE phase starts. Vertical lines mark various radii $R_{\rm L}$ corresponding to different orbital separations after CE ejection. The depletion in the $^{12}$C and $^{16}$O mass fractions at about $M_{\rm r}=0.7\rm M_{\odot}$ is inherited from the hydrogen burning during the main sequence.}
    \label{fig:core}
\end{figure}
\begin{figure*}
\centering
	\includegraphics[width=\textwidth]{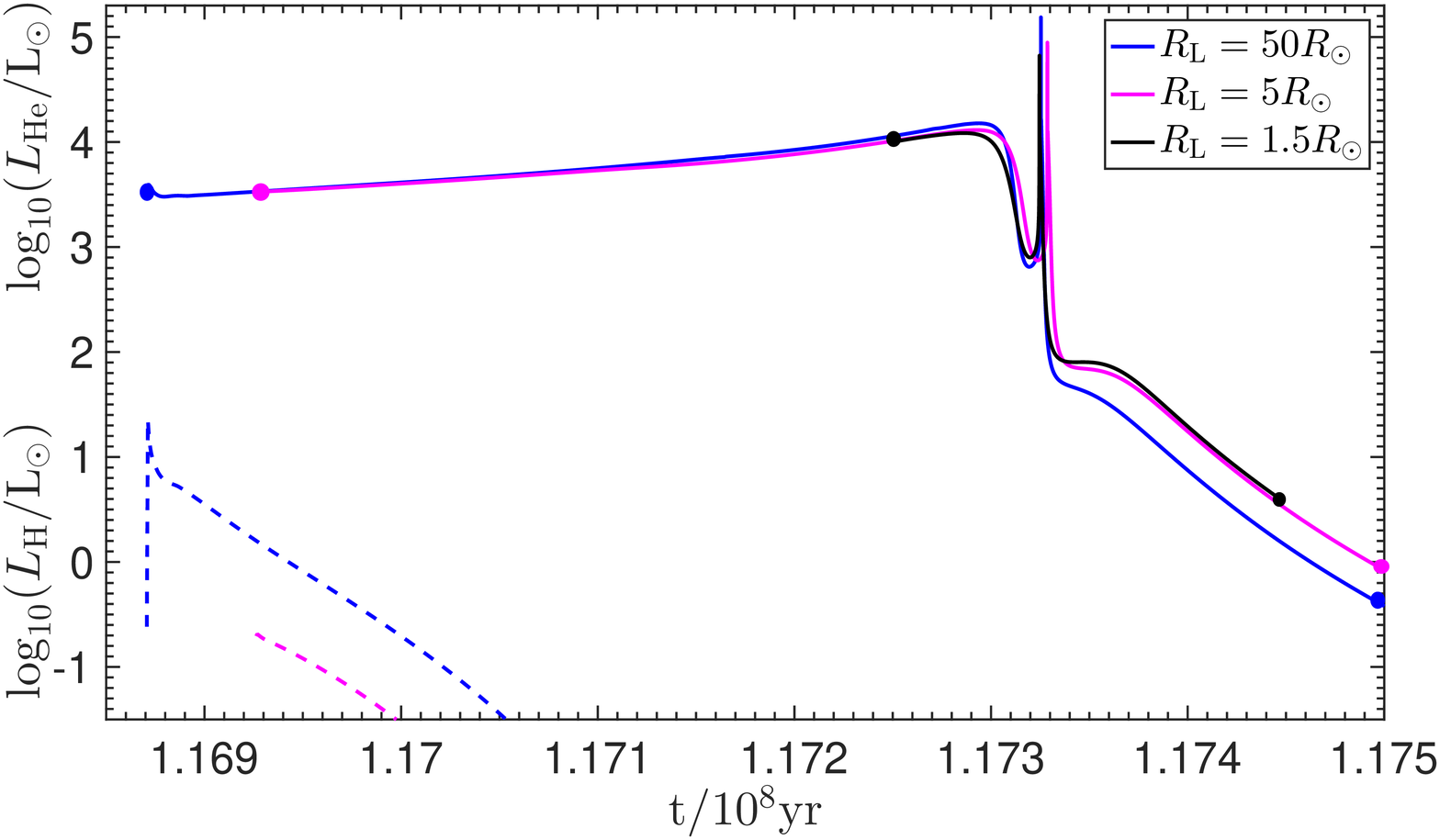}
    \caption{The hydrogen and helium luminosities shown with dashed and solid lines, respectively, during the post-CE RLOF phase as a function of evolutionary time for different  $R_{\rm L}$, or post-CE orbital separations. The beginning and end of the post-CE RLOF of each of the sequences are marked by dots of the same colour. For $R_{\rm L}  = 1.5\,\rm R_{\odot}$ there is no hydrogen shell so $L_{\rm H}=0$.}
    \label{fig:LH-He}
\end{figure*}
%
\subsection{Orbital separation after CE}\label{ssec:comp}
 The results above are for $R_{\rm L}= 24.5\,\rm R_{\odot}$. However, this is an arbitrary choice that we do not attempt to constrain by theory. This is because of the uncertainties enshrouding the CE evolution of the system during the complex hydrodynamical spiral-in, as well as the efficiency of the energy conversion to whatever is driving the envelope loss. To see how this choice affects the stripping of the primary and the final fate of the system, we assume different orbital separations at the end of the CE or, equivalently, different $R_{\rm L}$. We investigate the evolutionary behaviour when $R_{\rm L}\in \{1, 1.5, 1.9, 2, 3, 5, 33, 50\}\,\rm R_{\odot}$, as indicated relative to the core composition in Fig.~\ref{fig:core}. At the end of the CE phase, we apply the conditions given in Section \ref{ssec:RLOF} for stable mass transfer by RLOF. I.e. we choose the mass of a companion such that the mass ratio ensures stable mass transfer, calculating the binary separation from the Roche lobe radius of the primary, and finding the period using Kepler's third law. Table \ref{tab:table1} summarizes the properties of the post-CE systems and the final white dwarfs. We find that the evolution during the CE and RLOF in all model sequences with $R_{\rm L} \geq 1.9\,\rm R_{\odot}$ is similar and the post-CE remnants re-fill their Roche lobes. However, the post-CE remnants with $R_{\rm L} = 1$ and 1.5 $\rm R_{\odot}$ do not expand and thus fail to re-fill their Roche lobes after the CE phase. This is because in these two cases all the hydrogen envelope and most of the helium shell are already stripped in the CE phase as seen in Table \ref{tab:table1}.\\
\indent When $R_{\rm L}=1.9\,\rm R_{\odot}$ all the hydrogen envelope is stripped but not the helium shell. Because the star still re-fills its Roche lobe, we are sure that it is shell helium burning that drives the expansion and the subsequent RLOF following CE ejection.\\
\indent Fig.~\ref{fig:LH-He} shows the hydrogen and helium luminosities during the post-CE RLOF phase when $R_{\rm L}=1.5\,,\,5$ and $50\,\rm R_{\odot}$. We find that in all three cases the hydrogen luminosity is negligible. The spike in the helium luminosity is due to a helium flash in a thin shell. This confirms the connection between the Roche-lobe filling stars in the post-CE RLOF phase and the activity in the helium shell. In all nine model sequences, the primary ends up as a white dwarf of about $0.8\,\rm M_{\odot}$. Note that, if our post-CE naked helium stars were in very wide binaries that avoid interaction and thus post-CE RLOF, they would expand as helium giants and reach AGB dimensions. Therefore, as long as the helium shell is not completely stripped after the CE phase and the binary is close enough to allow interaction, post-CE RLOF occurs.\\  
\indent \citet{NO01} find that the removal of the helium layer requires a binary with the right conditions so that the primary can re-fill its Roche lobe, otherwise it can only lose its helium layer in a stellar wind. We find that the removal of the hydrogen layer and subsequently the helium layer is possible through a binary interaction resulting in double mass transfer. This is not sensitive to the post-CE orbital separation as long as the helium shell is not completely stripped during the CE phase and post-CE RLOF begins. We do not need any fine tuning of the binary parameters to strip both the hydrogen shell and most of the helium shell.
\section{Conclusions}\label{sec:conc}
\label{sec:conc}
We consider a binary system with a relatively long orbital period such that the more massive companion fills its Roche lobe on the EAGB. We strip the star by applying fast mass loss to mimic a CE event. After the CE phase, the donor is stripped of most of its envelope and has a thin hydrogen shell on the surface. When the system detaches, we allow the donor to refill its Roche lobe and undergo stable post-CE RLOF driven by shell helium burning. We find this phase to be prolonged and the core grows as the helium shell burns. By the end of the post-CE RLOF phase the donor is stripped of most of its helium shell and ends up as white dwarf of mass about $0.8\,\rm M_{\odot}$. We studied the sensitivity of our results to system parameters such as the mass of the companion and the pre-CE orbital separation. We find that the variation in the companion mass can change the final binary separation from a few days to about 100 d. When we vary the post-CE orbital separation we find that the donor refills its Roche lobe in the post-CE RLOF phase except in the cases when all the helium shell has already been stripped in the CE phase. Roche-lobe overflow in the post-CE RLOF phase is thus due to the burning in the helium shell. We find that no fine tuning of the binary system is required for the binary interaction to remove both the helium and hydrogen layers in such a double mass transfer mechanism, leaving all such systems with a similar $0.8\,\rm M_{\odot}$ CO white dwarf. 
\section*{Acknowledgements}
GMH and RGI thank the STFC for funding Rutherford grant ST/M003892/1. GMH thanks Wolfson College, Cambridge, for her fellowship and for using their library and facilities. RGI thanks the STFC for funding his Rutherford fellowship under grant ST/L003910/1 and Caracoli, Guildford for their lovely reading spot. CAT thanks Churchill College for his fellowship.

\bsp	
\label{lastpage}
\end{document}